\documentclass[12pt]{article}
\usepackage{amsfonts}
\usepackage{graphicx}
\usepackage{amsmath}
\usepackage{float}
\textwidth=6.5in \hoffset=-.75in \voffset=-1in \numberwithin{equation}{section}
\numberwithin{figure}{section}

\newcommand {\nn}{\nonumber}
\newcommand {\be}{\begin{equation}}
\newcommand {\ee}{\end{equation}}
\newcommand {\bea}{\begin{eqnarray}}
\newcommand {\eea}{\end{eqnarray}}

\begin{document}

\begin{titlepage}
\vspace{1cm}
\begin{center}
{\Large \bf {New Class of Exact Solutions in Einstein-Maxwell-dilaton Theory}}\\
\end{center}
\vspace{2cm}
\begin{center}
A. M. Ghezelbash{ \footnote{ E-Mail: amg142@campus.usask.ca}}
\\
Department of Physics and Engineering Physics, \\ University of Saskatchewan, \\
Saskatoon, Saskatchewan S7N 5E2, Canada\\
\vspace{1cm}
\vspace{2cm}

\end{center}

\begin{abstract}

We find new solutions to the five-dimensional Einstein-Maxwell-dilaton theory with cosmological constant where the dilaton field couples to the electromagnetic field as well as to the cosmological term with two different coupling constants.  The five-dimensional spacetime is non-stationary and is a conformally regular spacetime, everywhere.  Both the dilaton field and the electromagnetic field depends on time and two spatial directions. The cosmological constant takes positive, negative or zero value, depending on the value of coupling constant.  We study the physical properties of the spacetime and show that the solutions are unique in five dimensions and can't be uplifted to higher-dimensional Einstein-Maxwell theory or Einstein gravity in presence of cosmological constant. Moreover, we construct new solutions to the theory where both coupling constants are equal.

\end{abstract}
\end{titlepage}\onecolumn 
\bigskip 

\section{Introduction}

One of the main objectives in gravitational physics is to construct, explore and understand the exact solutions to the Einstein field equations in the background of matter fields in different dimensions. 
The possibility of extending the known solutions in asymptotically flat spacetime to the asymptotically de-Sitter and anti de-Sitter solutions, is an important step in better understanding the recent holographic proposals between the extended theories of gravity and the conformal field theories in different dimensions \cite{hol1}-\cite{hol2}. The explored solutions cover a vast area of solutions with different charges, such as  NUT charges \cite{awad}-\cite{holend}, different matter fields, such as Maxwell field, dilaton field \cite{TNme,T1} and axion field \cite{T2} and black hole solutions with different horizon topologies \cite{Ch1}-\cite{Jap}.  Moreover, the references \cite{BL1}-\cite{BL4} include the other solutions to extended theories of gravity  with different type of matter fields in different dimensions.  The class of solutions to Einstein-Maxwell-dilaton theory, in which the dilaton field interacts with the cosmological constant and the Maxwell field, was considered in \cite{Torii}-\cite{Kino}. These solutions are relevant to the generalization of the Freund-Rubin compactification of M-theory \cite{FRC}, \cite{FRC2}. Inspired by construction and exploring the new exact solutions to the field equations of Einstein-Maxwell theory, in this article, we find exact analytical solutions to the Einstein-Maxwell-dilaton theory with cosmological constant and two dilaton coupling constants. 

We explicitly consider the Einstein-Maxwell-dilaon theory in five dimensions, where the dilaton field couples to the Maxwell field as well as the cosmological term with two different coupling constants. We solve and show that the field equations support a non-stationary spacetime with non-trivial solutions for the dilaton and Maxwell fields.  Moreover, we consider the case where the two coupling constants are equal and non-zero and find a new class of exact solutions. We discuss the physical properties of the solutions. Finally, we consider the theory where both the coupling constants are zero and show that there is a class of non-trivial solutions to the field equations.  We also show that our new exact solutions can not simply be found as the result of compactification of a higher (than five) dimensional Einstein or Einstein-Maxwell theory with a cosmological constant;
however we show that our new class of solutions can be uplifted to a higher-dimensional gravity theory with a form-field, without a cosmological constant. 

In section \ref{sec:aneqb}, we consider the Einstein-Maxwell-dilaton theory in presence of the cosmological constant with two different dilaton coupling constants. We consider the self-dual Eguchi-Hanson space as the spatial section of the five-dimensional spacetime and also consider special ansatze for the Maxwell field and the dilaton field. The spacetime is described by two metric functions, where one of them depends explicitly on time and the other depends on radial and angular coordinates. We solve the field equations  and find that the dilaton coupling constants must be related to each other. Moreover, we show that the cosmological constant is related to the dilaton coupling constant and becomes negative, positive or zero, depending on dilaton coupling constant. We then explicitly show that it is not possible to uplift the solutions into a higher than five dimensional Einstein-Maxwell theory with a cosmological constant or into a six-dimensional gravity with a cosmological constant. In section \ref{sec:aeqb}, we consider the Einstein-Maxwell-dilaton theory, where the two dilaton coupling constants are equal to each other. We employ a metric ansatz that is very similar to the metric ansatz where the dilaton coupling constants are not equal.  In the former case, we consider an extra time dependence for the metric function, that previously depended only on radial and angular coordinates. We also consider a different form for the dependence of the Maxwell field on the metric function. 
We explicitly solve the field equations and find exact solutions for the metric functions, the Maxwell field and the dilaton filed. Moreover, we find that the cosmological constant can take only some specific values in terms of the dilaton coupling constant. We discuss the physical properties of the solutions and compare the asymptotic behaviour of the solutions to the case, where the dilaton coupling constants are not equal. We also discuss the possibility of uplifting the solutions into higher-dimensional Einstein-Maxwell theory, only for some specific values of the dilaton coupling constant. 

Finally, in section \ref{sec:aeq0}, we consider a very special case where the dilaton coupling constant is zero. The dilaton sector decouples from the theory and the theory reduces to the Einstein-Maxwell theory.  We consider the metric ansatz and the gauge field the same as in section \ref{sec:aeqb}, where we set the dilaton constant to be zero. We find exact analytical solutions for the metric functions. 
The concluding remarks and two appendices, wrap up the article.

\section{Exact Solutions to Einstein-Maxwell-dilaton field equations with two different coupling constants $a$ and $b$}
\label{sec:aneqb}

The Eguchi-Hanson space is an asymptotically locally Euclidean space where its curvature two form is a self-dual form. The Eguchi-Hanson line element is given by
\be
ds_{EH}^2=\frac{dr^2}{g(r)^2}+\frac{r^2g(r)^2}{4}(d\psi +\cos \theta d \phi)^2+\frac{r^2}{4}(d\theta^2+\sin^2(\theta)d\phi^2).\label{EH}
\ee
The metric function $g(r)$ in (\ref{EH}) is given by $g(r)=\sqrt{1-(\frac{h}{r})^4}$ that implies the radial coordinate  $r \geq h$. 
The Eguchi-Hanson metric (\ref{EH}) can be expressed in terms of three one-forms $\sigma_i,i=1,2,3$ as
\be
ds_{EH}^2=\frac{dr^2}{g(r)^2}+\frac{r^2g(r)^2}{4}\sigma_3^2+\frac{r^2}{4}(\sigma_1^2+\sigma_2^2),\label{EH2}
\ee
where the one-forms $\sigma_i$, are the left invariants for the $SU(2)$ group manifold which are 
\begin{eqnarray}
\sigma_1&=&\sin \psi d\theta-\cos \psi\sin \theta d \phi,\\
\sigma_2&=&-\cos \psi d\theta-\sin \psi\sin \theta d \phi,\\
\sigma_3&=&d\psi +\cos \theta d \phi.
\end{eqnarray}
The three periodic Euler angles $\theta, \phi$ and $\psi$ parametrize the $SU(2)$ group manifold, where their periodicities are  $\pi, 2\pi$ and $4\pi$, respectively. We note that there is a single removable bolt singularity at $r=h$ that can be removed by restricting the range of coordinate $\psi$ in the interval $0\leq \psi \leq 2\pi$ for any positive Eguchi-Hanson parameter $h$. The Kretschmann invariant for the Ricci-flat Eguchi-Hanson metric (\ref{EH}) is
\be
{\cal K}=\frac{384h^8}{r^{12}}.\label{KRE}
\ee
The topology of Eguchi-Hanson space near the bolt singularity at $r=h$, is the topology of manifold $R^2\times S^2$ where the radius of $S^2$ is $\frac{h}{2}$.  Moreover,  the Eguchi-Hanson space is asymptotically a lens space $S^3/\mathbb{Z}_2$.  The supergravity solutions based on Eguchi-Hanson space, have  been constructed and investigated in \cite{6dcone, M2}. The CFT realization of the heterotic strings in the double scaling limit, has been constructed by using the warped extension of Eguchi-Hanson space \cite{warp}.  The coalescence of black holes, as well as black ring and brane solutions on the Eguchi-Hanson space were extensively studied in \cite{Co2}-\cite{BR3}. 
Moreover, the six-dimensional Eguchi-Hanson space was obtained in \cite{Tse} as a Ricci-flat K{\" a}hler metric on a resolved conifold and used for construction of M-brane solutions on resolved conifolds \cite{6dcone}. 

The action for the five-dimensional Einstein-Maxwell-dilaton theory, where the dilaton field couples  to the electromagnetic field, and also couples to the cosmological constant, is given by
\begin{equation}
S=\int d^5x \sqrt{-g}\{ R-\frac{4}{3}(\nabla \phi)^2-e^{-4/3a\phi}F^2-e^{4/3b\phi}\Lambda\},
\label{act}
\end{equation}
where we set the gravitational constant equal to $(16\pi)^{-1}$. The coupling constant for the coupling of the dilaton field to the electromagnetic field is $a$. Moreover,  $b$ is the coupling constant for the coupling of the dilaton field to the cosmological term.  We obtain the Einstein's field equations ${\cal E}_{\mu\nu}=0$, together with the field equations ${\cal M}_{\mu}=0$ for the electromagnetic field and the field equation ${\cal D}=0$ for the dilaton field, by varying the action (\ref{act}) with respect to the metric tensor $g_{\mu\nu}$, the gauge field ${A_\mu}$ and the dilaton field $\phi$, respectively. The field equations ${\cal E}_{\mu\nu}={\cal M}_{\mu}= {\cal D}=0$ are given explicitly in appendix A.

\begin{figure}[H]
\centering
\includegraphics[width=0.4\textwidth]{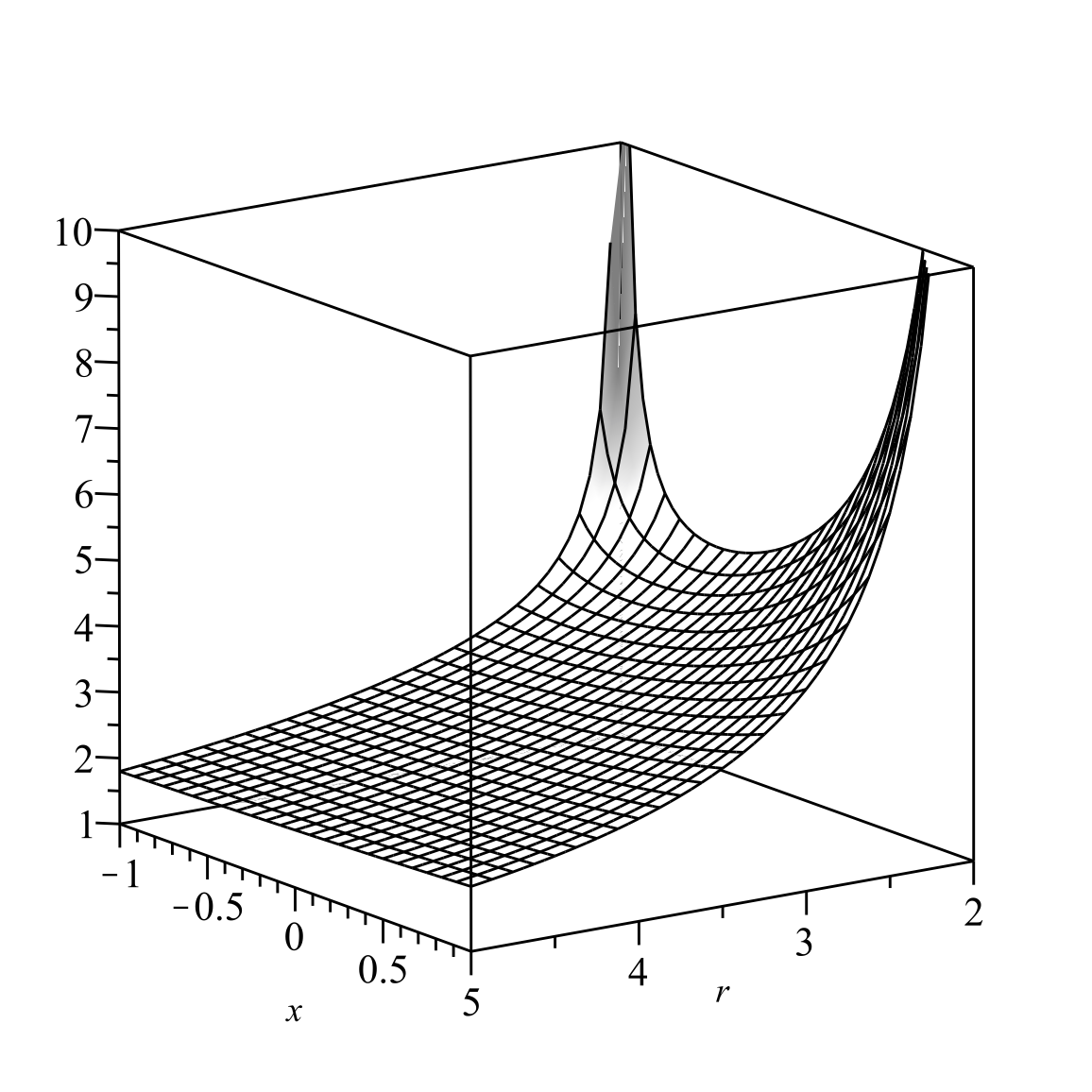}

\caption{The metric function $H(r,\theta)$ as function of $r$ and $x=\cos\theta$, where we set $h=2,\,g_+=5,\,g_-=3,\,a=1$.}
\label{fig1}
\end{figure}

In this section, we consider the most general case for the coupling constants, in which the coupling constants are non-zero numbers and are not equal to each other.  In this regard, we consider the five-dimensional metric and the electromagnetic gauge field as
\begin{equation}
ds_5^{2}=-\frac{1}{H^2(r,\theta)}dt^{2}+R^2(t)H(r,\theta)ds_{EH}^2,
\label{dsanoteqb}
\end{equation}
\begin{equation}
{A_t}(t,r,\theta)={\alpha R^X(t)}{H^Y(r,\theta)}\label{gaugeanoteqb},
\end{equation}
where the metric functions $H(r,\theta)$ and $R(t)$, depend on two coordinates, $r,\,\theta$ and $t$, respectively. We also note that the gauge field (\ref{gaugeanoteqb}) depends on three coordinate $t,\,r$ and $\theta$ where $X$ and $Y$ are two constants. The gauge field (\ref{gaugeanoteqb}) generates an electric field, in $r$ and $\theta$ directions, that depends on coordinates $t,\,r$ and $\theta$.  In equation (\ref{dsanoteqb}), $ds_{EH}^2$ is the Eguchi-Hanson space which is given by equation (\ref{EH}). Moreover, we consider the following ansatz for the dependence of the dilaton field on metric functions  $H(r,\theta)$ and $R(t)$
\be
\phi(t,r,\theta)=-\frac{3}{4a}\ln(H^U(r,\theta)R^V(t))\label{dilatonanoteqb},
\ee
where $U$ and $V$ are two other constants. We should emphasize that the class of our solutions to Einstein-Maxwell-dilaton theory (that are given in equations (\ref{dsanoteqb}), (\ref{gaugeanoteqb}) and (\ref{dilatonanoteqb})), depends on time coordinates as well as two spatial coordinates $r$ and $\theta$ and to our knowledge, is the first know class of solutions depending on three coordinates.  The class of our solutions is quite distinct from the other known classes of solutions \cite{TNme,chris2}. 
\newline
The two non-zero Maxwell's equations ${\cal M}^r=0,\, {\cal M}^\theta=0$ yield 
\be
X+V=-2. \label{eq1}
\ee
The only other non-zero Maxwell's equation ${\cal M}^{t}=0$ gives a coupled differential equation for the metric function $H(r,\theta)$ as
\begin{eqnarray}
&&(UY+1) (-h^4+r^4) r \sin \theta  \left( {\frac {\partial H}{\partial r}} \right) ^{2}
-  ( {{h}}^{4}r-r^5)
H  \sin  \theta {
\frac {\partial ^{2}H}{\partial {r}^{2}}}  
+ ( {{h}}^{4}+3r^4) {
H\sin \theta \frac {\partial H}{\partial r}}  
 \nn\\
&+&4(U+Y+1)
\sin \theta  \left( {\frac {\partial H}{\partial \theta}
} \right) ^{2}{r}^{3}
+4H  \sin  \theta {\frac {\partial ^{2} H}{
\partial {\theta}^{2}}}  {r}^{3}
+4H\cos  \theta 
  {\frac {\partial H}{\partial \theta}}
   {r}^{3}
 =0.
\label{Mt}
\end{eqnarray}
Moreover the non-diagonal Einstein's equations ${\cal G}_{tr}=0,\, {\cal G}_{t\theta}=0$ implies 
\be
UV+4a^2=0.\label{eq2}
\ee
The other non-diagonal Einstein's equation ${\cal G}_{r\theta}=0$ implies 
\begin{equation}
U+2Y=0,\,V+2X=0,\,\alpha^2Y^2=\frac{3U^2}{4a^2}+\frac{3}{2}.\label{eq3}
\end{equation}
The equations (\ref{eq1}),(\ref{eq2}) and (\ref{eq3}) yield
\be
X=2,\,Y=-1-\frac{a^2}{2},\,U=a^2,\,V=-4,\,\alpha^2=\frac{3}{a^2+2}.\label{numvalues}
\ee
Upon substituting the results from equations (\ref{numvalues}) and (\ref{dilatonanoteqb}) in (\ref{Mt}), we find a partial differential equation for $H(r,\theta)$. The form of this equation suggests that we change the function $H(r,\theta)$ to $G(r,\theta)=H(r,\theta)^{1+\frac{a^2}{2}}$. The partial differential equation for $G(r,\theta)$ read as
\be
(h^4r-r^5) \sin \theta {\frac {\partial ^{2}G}{\partial {r}
^{2}}} - (h^4+3r^4)\sin\theta {\frac {\partial G}{
\partial r}}-4r^3(\sin  
\theta {\frac {\partial ^{2}G}{\partial {\theta}^{2}}}+\cos  \theta
 {\frac {\partial G}{\partial \theta}})=0.
\label{SS4}
\ee
We find that the solutions to differential equation (\ref{SS4}) are proportional to $\frac{1}{r^2\pm h^2\cos\theta}$ and so we consider a general solution for $G$, such as 
\be
G(r,\theta)=1+\frac{g_+}{r^2+ h^2\cos\theta}+\frac{g_-}{r^2- h^2\cos\theta},\label{GG}
\ee
where $g_{\pm}$ are two constants. It turns out that the metric function $H(r,\theta)$ is then equal to
\be
H(r,\theta)=(1+\frac{g_+}{r^2+ h^2\cos\theta}+\frac{g_-}{r^2- h^2\cos\theta})^{\frac{2}{2+a^2}}.
\ee
Figure \ref{fig1} shows the behaviour of metric function $H(r,\theta)$ where we set $h=2,\,g_+=5,\,g_-=3,\,a=1$.

We then substitute all the results into equation ${\cal G}_{tt}=0$ and solve for $\Lambda$ in terms of unknown metric function $R(t)$. We substitute the result for $\Lambda$ in equation ${\cal G}_{rr}=0$ and find a differential equation for $R(t)$ that is given by
\be
a^2R(t)\frac{d^2R(t)}{dt^2}=(a^2-4)(\frac{dR(t)}{dt})^2.\label{eqforR}
\ee
The solutions to differentia equation (\ref{eqforR}) are
\be
R(t)=(\xi\,t+\zeta)^{a^2/4},\label{RRR}
\ee
where $\xi$ and $\zeta$ are two arbitrary constants.
If we substitute for $R(t)$ (given by (\ref{RRR})) in the equation that we previously found for $\Lambda$, we find that the cosmological constant is  given by
\be
\Lambda=\frac{3}{8}Q(r,\theta)^{\frac{2ab+4}{a^2+2}}(\xi\,t+\zeta)^{-ab-2}a^2\xi^2(a^2-1),\label{LamQ}
\ee
where $Q(r,\theta)$ is a function of $r$ and $\theta$. The presence of $Q$ as a function of coordinates in (\ref{LamQ}) enforces a relation between the coupling constants,  $ab=-2$, to guarantee that the cosmological constant is independent of spacetime coordinates. Hence we find that the cosmological constant is related to the coupling constant $a$ and $\xi$ by
\be
\Lambda=\frac{3}{8}\xi^2a^2(a^2-1).\label{cosmo}
\ee
We note that the cosmological constant takes a negative value for  the coupling constant $0<a<1$, positive value for $a>1$ and zero where $a=1$.

\begin{figure}[H]
\centering
\includegraphics[width=0.4\textwidth]{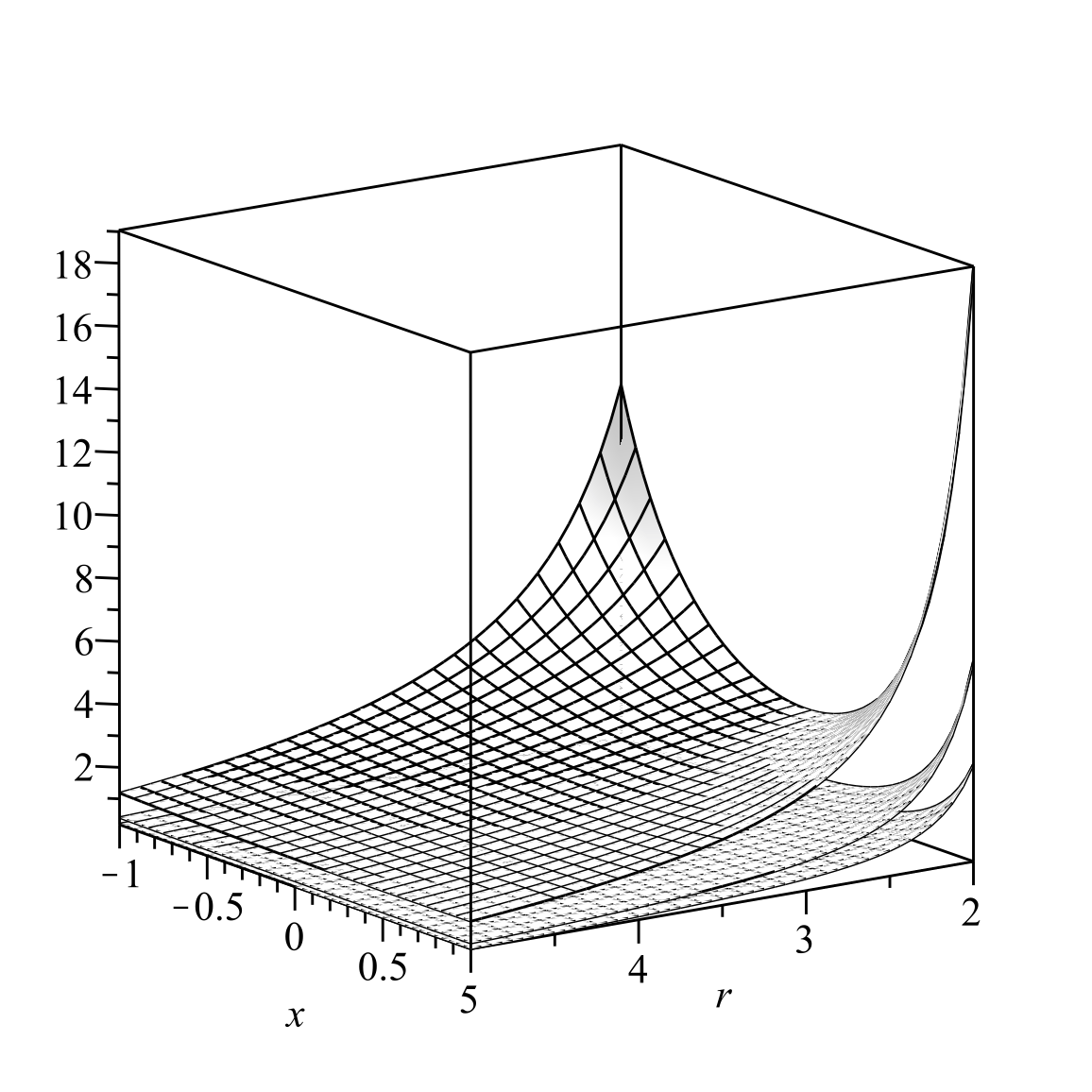}

\caption{The $r$-component of electric field as function of $r$ and $x=\cos\theta$ for three time slices; $t=1$ (upper surface), $t=10$ (middle surface) and $t=100$ (lower surface), where we set $h=2,\,g_+=5,\,g_-=3,\,a=1,\,\xi=2,\,\zeta=4$.}
\label{fig2}
\end{figure}

\begin{figure}[H]
\centering
\includegraphics[width=0.4\textwidth]{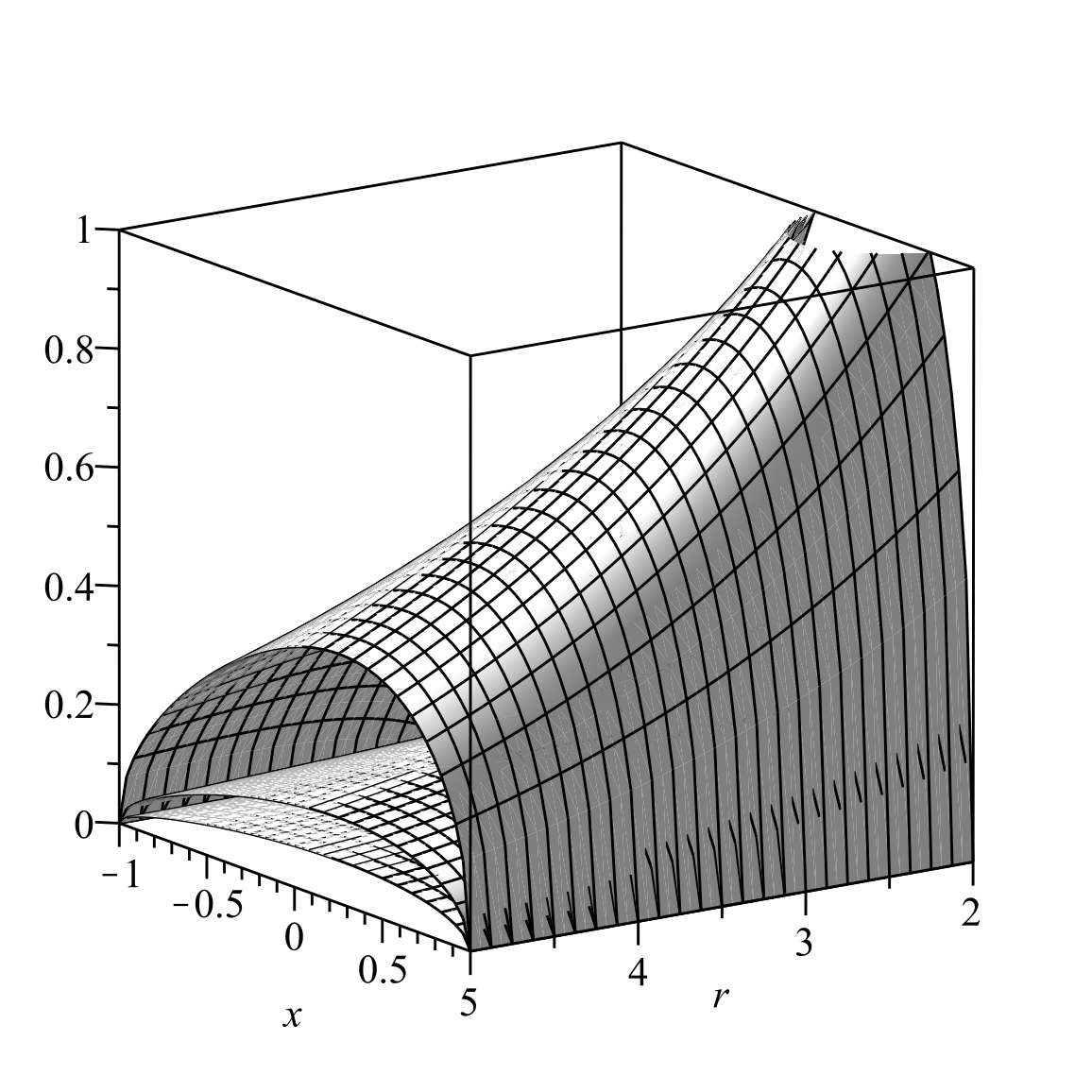}

\caption{The $\theta$-component of electric field 
as function of $r$ and $x=\cos\theta$ for three time slices; $t=1$ (upper surface), $t=10$ (middle surface) and $t=100$ (lower surface), where we set $h=2,\,g_+=5,\,g_-=3,\,a=1,\,\xi=2,\,\zeta=4$.}
\label{fig3}
\end{figure}

Upon substitution of cosmological constant $\Lambda$ and the coupling constant $b=-\frac{2}{a}$ in the other Einstein's equations and also the dilaton equation, we find that all the equations are completely satisfied.  We should emphasize that the coupling constant $a$ is completely arbitrary in our solutions, while the coupling constant $b$ is related to $a$ by $b=-\frac{2}{a}$.  In figures \ref{fig2} and \ref{fig3}, we present the components of the electric field in $r$ and $\theta$ directions as functions of $r$ and $\theta$, for three different time slices. 

\begin{figure}[H]
\centering
\includegraphics[width=0.4\textwidth]{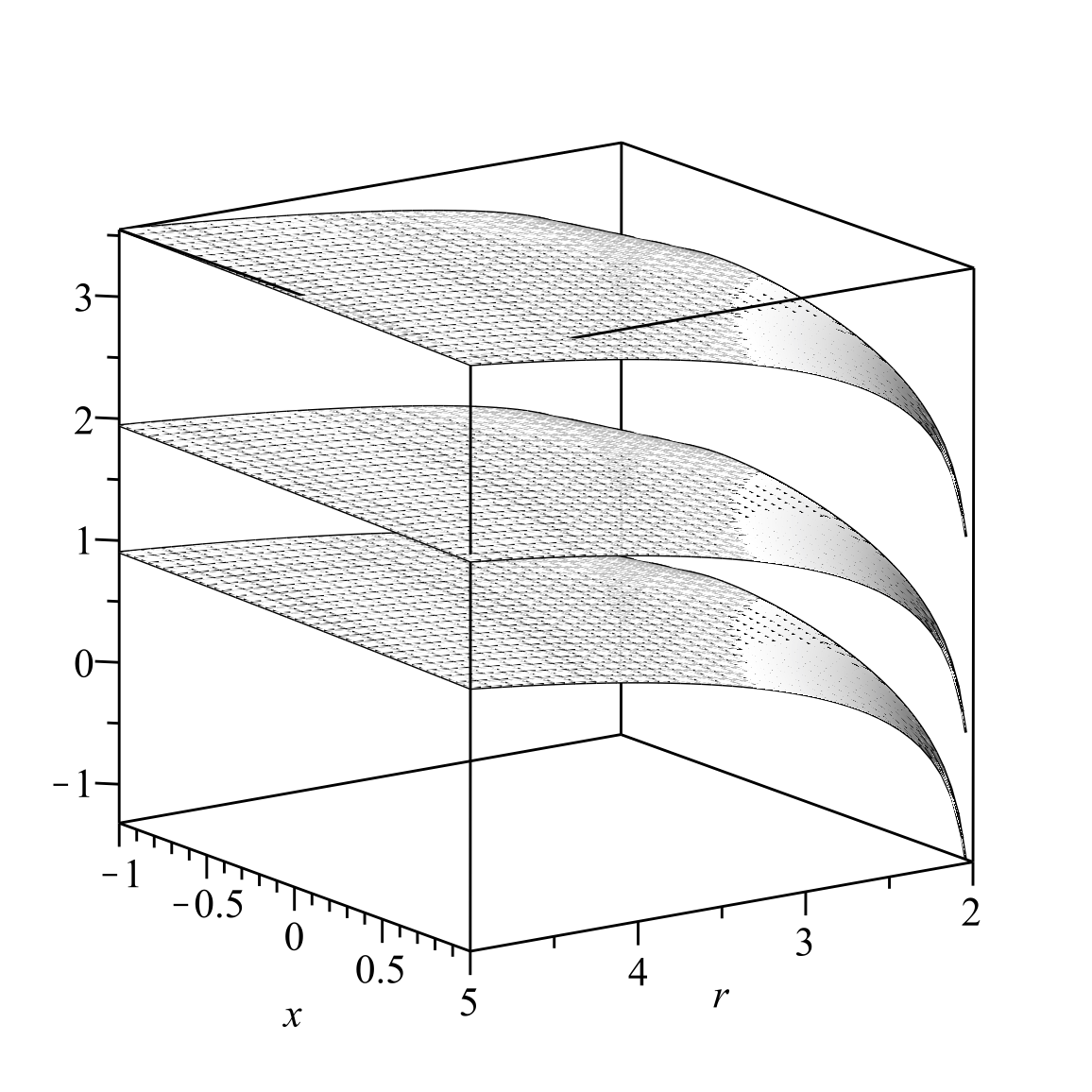}

\caption{The dilaton field $\phi(t,r,\theta)$ as function of $r$ and $x=\cos\theta$ for three time slices; $t=1$ (upper surface), $t=10$ (middle surface) and $t=100$ (lower surface), where we set $h=2,\,g_+=5,\,g_-=3,\,a=1,\,\xi=2,\,\zeta=4$.}
\label{fig4}
\end{figure}

Moreover, figure \ref{fig4} shows the behaviour of the dilaton field, as a function of $r$ and $\theta$ for three different time slices. 

Furnished by all the above results, we find that the five-dimensional spacetime (\ref{dsanoteqb}) read explicitly as
\begin{eqnarray}
ds_5^2&=& -\{\frac{r^4+g_1r^2-g_2h^2\cos\theta-h^4\cos^4\theta}{r^4-h^4\cos^2\theta}\}^{-4/(a^2+2)}dt^2\nn\\
&+&(\xi t+\zeta)^{a^2/2}\{\frac{r^4+g_1r^2-g_2h^2\cos\theta-h^4\cos^4\theta}{r^4-h^4\cos^2\theta}\}^{2/(a^2+2)}\nn\\
&\times&\{\frac{dr^2}{g(r)^2}+\frac{r^2g(r)^2}{4}(d\psi +\cos \theta d \phi)^2+\frac{r^2}{4}(d\theta^2+\sin^2(\theta)d\phi^2)\},\label{metric5danoteqb}
\end{eqnarray}
where $g_1=g_++g_-$ and $g_2=g_+-g_-$. 

We note that the metric (\ref{metric5danoteqb}) possesses a bolt structure at $r=h$ and there are two singularities on north and south poles ($\theta=0$ and $\theta=\pi$, respectively) on $S^2$ bolt.  Moreover, the Ricci scalar is singular on the hypersurface $t=-\frac{\zeta}{\xi}$.  The bolt can be removed by restricting the range of coordinate $\psi$ on the interval $[0,2\pi]$. Moreover, the spacetime becomes completely regular on the hypersurface  $t=-\frac{\zeta}{\xi}$, by choosing the dilaton coupling constant $a \geq \sqrt{2}$, after conformally rescaling the metric by the factor $e^{-\frac{4}{3}a\phi}$.
We may consider that the solution for the spacetime metric (\ref{metric5danoteqb}) along with the  electromagnetic field (\ref{gaugeanoteqb}) and the dilaton (\ref{dilatonanoteqb}), can be uplifted to some known solutions in higher dimensions with a cosmological constant. However, we could not find any set of coupling constants such that our solutions can be uplifted into a higher-dimensional theory. More specifically, there are two different possible uplifting processes.  In the first uplifting process, some solutions to the Einstein-Maxwell-dilaton theory (with two coupling constants and the cosmological term, as the same as in action (\ref{act})), can be uplifted to the Einstein-Maxwell theory with cosmological constant in higher than five dimensions. In the second uplifting process, the solutions can be uplifted to the solutions of six dimensional gravity with a cosmological constant.   

However, none of these two uplifting processes are applicable to our solutions (\ref{metric5danoteqb}), (\ref{gaugeanoteqb}) and (\ref{dilatonanoteqb}) to the Einstein-Maxwell-dilaton theory. In fact, the former uplifting from five to $5+d$-dimensions is possible only for some specific values of the coupling constants where $a=b$ and the number of extra dimensions is equal to $d=\frac{3a^2}{1-a^2}$ \cite{GK}, \cite{GSSST}. However as we noticed before, the consistency of the solutions implies $ab=-2$ and so, there is no value for the coupling constant $a$, such that $a=b$.

The latter uplifting from five to six dimensions (i.e. the special case of uplifting the $d$-dimensional solutions to Einstein-Maxwell-dilaton theory with two coupling constants to the solutions of $d+1$-dimensional gravity with the cosmological constant) is only possible if the coupling constants are equal to $a=\pm 2$ and $b=\pm \frac{1}{2}$ , respectively for $d=5$ \cite{chris2}, \cite{Steer}.  However, the product of these coupling constants are equal to $+1$ and so they are not in agreement with the relation between our coupling constants.

Indeed, we have explicitly checked that the six-dimensional metric for the uplifted theory with $b=\pm\frac{1}{2}$ \cite{chris2}, \cite{Steer}
\begin{equation}
ds_6^2=e^{\mp(\frac{4}{3})(\frac{1}{2})\phi(t,r,\theta)}ds_5^2+e^{\pm4(\frac{1}{2})\phi(t,r,\theta)}(dz+2A_t(t,r,\theta)dt)^2,\label{upl}
\end{equation}
does not satisfy the Einstein's equations with a cosmological constant. In (\ref{upl}), $z$ is the uplifted coordinate and $\phi(t,r,\theta)$ and $A_t(t,r,\theta)$ are given by equations (\ref{dilatonanoteqb}) and (\ref{gaugeanoteqb}), respectively.  A more careful analysis of the uplifting process for the different cases in references \cite{TNme, Steer} show that the  spacetime metric for the five-dimensional Einstein-Maxwell-dilaton theory, always is diagonal \cite{TNme, Steer}. Although there is no restriction on the diagonality of five-dimensional spacetime,
 but we may need 
a new anstaz for the uplifting of the five-dimensional spacetime (\ref{metric5danoteqb}) into the six-dimensional gravity with the cosmological constant. 

As we noticed above, the solution for the spacetime metric (\ref{metric5danoteqb}), the  electromagnetic field (\ref{gaugeanoteqb}) and the dilaton field (\ref{dilatonanoteqb}), can't be  uplifted to some known solutions in higher dimensions with a cosmological constant. However, as it is shown in reference \cite{GK}, we can consider the $D=p+q+1$-dimensional theory for the gravity with a $q+1$-potential ${\cal B}_{[q+1]}$ in presence of cosmological constant $\Lambda_D$
\begin{equation}
S_{D}=\int d^Dx\sqrt{-g}\{R-\frac{1}{2(q+2)!}{\cal F}_{[q+2]}^2+2\Lambda_D\}\label{HighD},
\end{equation}
where ${\cal F}_{[q+2]}$ is the $q+2$-field strength form for the $q+1$-potential ${\cal B}_{[q+1]}$, given by ${\cal F}_{[q+2]}=d{\cal B}_{[q+1]}$. The dimensional reduction from $D$-dimensions to $p+1$-dimensions on an internal  curved $q$-dimensional space with the line element $dK^2_{q}$, according to the ansatze for the metric
\begin{equation}
ds^2_{D}=e^{-\delta \hat \phi}ds^2_{p+1}+e^{\hat \phi(\frac{2}{\delta (p-1)}-\delta)}dK^2_{q},
\end{equation}
and the potential
\begin{equation}
{\cal B}_{[q+1]}={\cal A}_{[1]}\wedge dK_{q},
\end{equation}
leads to the Einstein-Maxwell-dilaton theory with a potential that depends on two exponentials,
\begin{equation}
S_{p+1}=\int d^{p+1}x \{ R-\frac{1}{2}(\nabla \hat \phi)^2-\frac{1}{4}e^{\gamma \hat \phi}{\cal F}_{[2]}^2+2\Lambda_D e^{-\delta \hat \phi}+2\hat \Lambda e^{-\frac{2}{\delta(p-1)}\hat \phi}\}
\label{EMD2E}.\end{equation}
In equation (\ref{EMD2E}), the dilaton coupling constants $\delta=\sqrt{\frac{2q}{(p-1)(p+q-1)}}$ and $\gamma=(2-p)\delta$ \cite{GK}. Moreover $\hat \Lambda$ is equal to the half of the Ricci scalar of the internal space. Comparing the actions (\ref{EMD2E}) and (\ref{act}), we note that our five-dimensional solutions (\ref{metric5danoteqb}), (\ref{gaugeanoteqb}) and (\ref{dilatonanoteqb}) can be uplifted to a higher-dimensional theory (\ref{HighD}) without any cosmological constant $\Lambda_D=0$, if we redefine $\gamma=-\frac{4}{3}\sqrt{\frac{3}{8}}a$, $\delta=-\frac{3}{2}\sqrt{\frac{3}{8}}\frac{1}{(p-1)b}$. The relation $ab=-2$ together with $\gamma=(2-p)\delta$ lead to $p=4$. Moreover, we have redefined the dilaton field $\hat \phi=\sqrt{\frac{8}{3}}\phi$, the one form ${\cal A}_{[1]}=2A_{t}dt$ and $2\hat \Lambda=-\Lambda$ to make all the terms in the action (\ref{EMD2E}) exactly equal to the terms in (\ref{act}).
The Ricci scalar ${\cal R}$ of spacetime (\ref{dsanoteqb}) is given by
\begin{eqnarray}
(2V ^3 H ^3 R  ^2r^2){\cal R} 
&=&16 H ^{5}RV    ^{3}  {\ddot {R}}  {r}^{2}+24H ^{5} {\dot R} ^{2} V ^{3}{r}^{2}+2 V_{rr}  V   H  ^{2}{r}^{2}-4V_r  ^{2}  H  ^{2}{r}^{2}-2H_{rr}  H  V  ^{2}{r}^{2}
\nn\\
&+& 2V {}V_{r} {}H_{r}{r}^{2}- H_{r,r} ^{2} V  ^{2}{r}^{2}  +16 V ^{3} H ^{2}+14 V_{r} V  H  ^{2}r-8H_{\theta\theta}  V  ^{3}H
\nn\\
&-&8\frac{\cos  \theta }{\sin \theta}V ^{3} H_\theta  H-6   V ^{2}H   H_r   r-4H_\theta ^{2}V ^{3}-16V ^{2} H  ^{2},
\end{eqnarray}
where $H_i=\frac{\partial H}{\partial i},\, H_{ii}=\frac{\partial ^2 H}{\partial i^2},\, i=r,\theta$ and $\dot {} =\frac{d}{dt}$. The Ricci scalar and also the quadratic Kretschmann scalar are divergent at $r=h,\, \theta=0$ and $r=h,\, \theta=\pi$. However by rescaling the metric by a conformal factor proportional to the dilaton field and restricting the range of dilaton coupling constant, we find a regular spacetime.  In the asymptotic limit where $r\rightarrow \infty$, the metric (\ref{metric5danoteqb}) approaches to
\be
ds_5^2 =-dt^2+(\xi t+\zeta)^{a^2/2}\left(dr^2+\frac{r^2}{4}((d\psi+\cos\theta d\phi)^2+d\theta^2+\sin^2\theta d\phi)\right).\label{rinf}
\ee
We note that on a fixed time hypersurface $t=T$, the asymptotic geometry is the geometry of a circle of radius $\frac{r}{2}(\xi T+\zeta)^{a^2/4}$ fibered over a sphere. The radius of circle at $T\rightarrow \infty$ is going to infinity as $\frac{r}{2}(\xi T)^{a^2/4}$. 



\section{Exact Solutions with equal coupling constants $a$ and $b$}
\label{sec:aeqb}

\begin{figure}[H]
\centering
\includegraphics[width=0.4\textwidth]{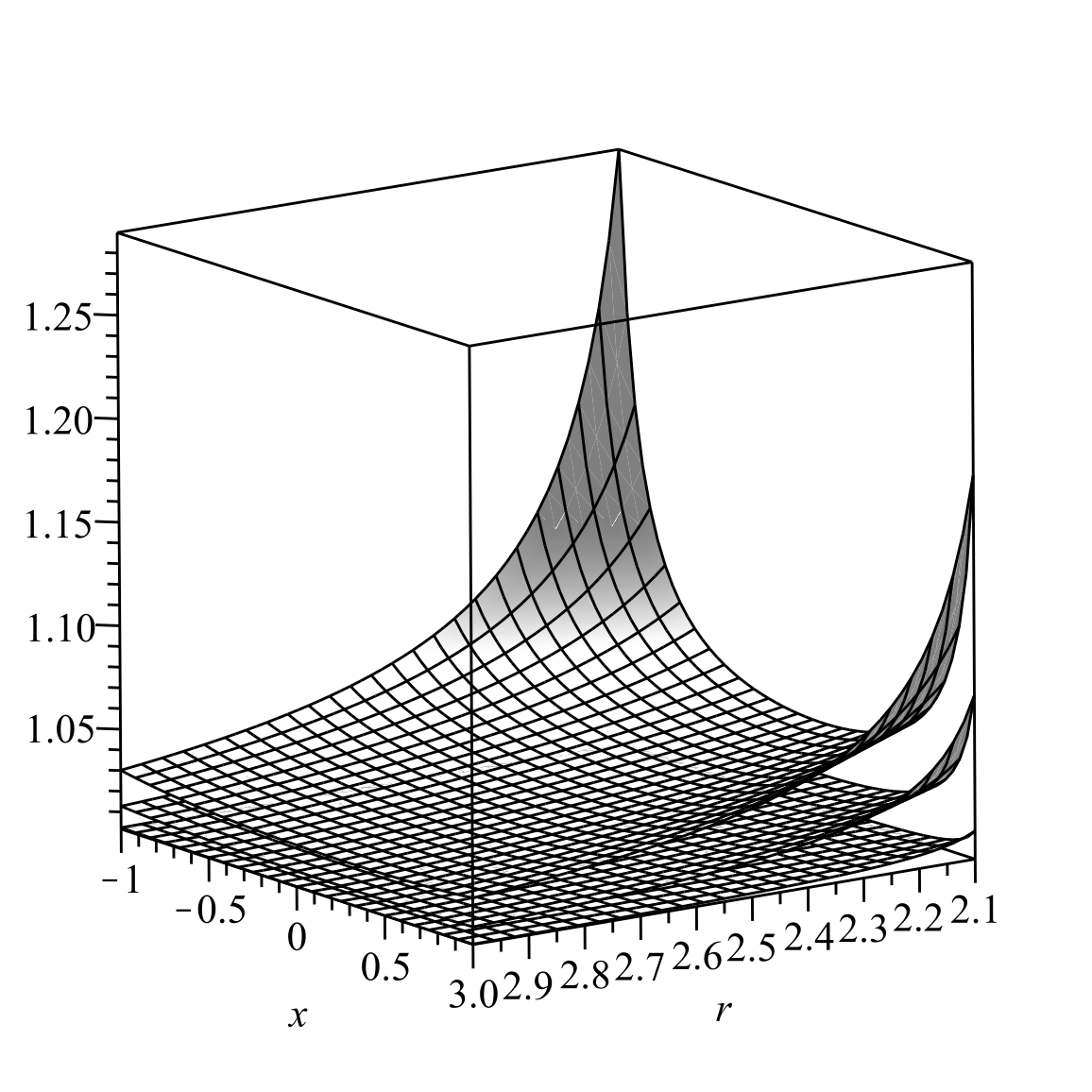}

\caption{The metric function $H(t,r,\theta)$ as function of $r$ and $x=\cos\theta$  for three time slices $t=1$ (upper surface), $t=2$ (middle surface) and $t=5$ (lower surface), where we set $h=2,\,k_+=5,\,k_-=3,\,a=1,\mu=1,\nu=2$.}
\label{fig5}
\end{figure}

\begin{figure}[H]
\centering
\includegraphics[width=0.4\textwidth]{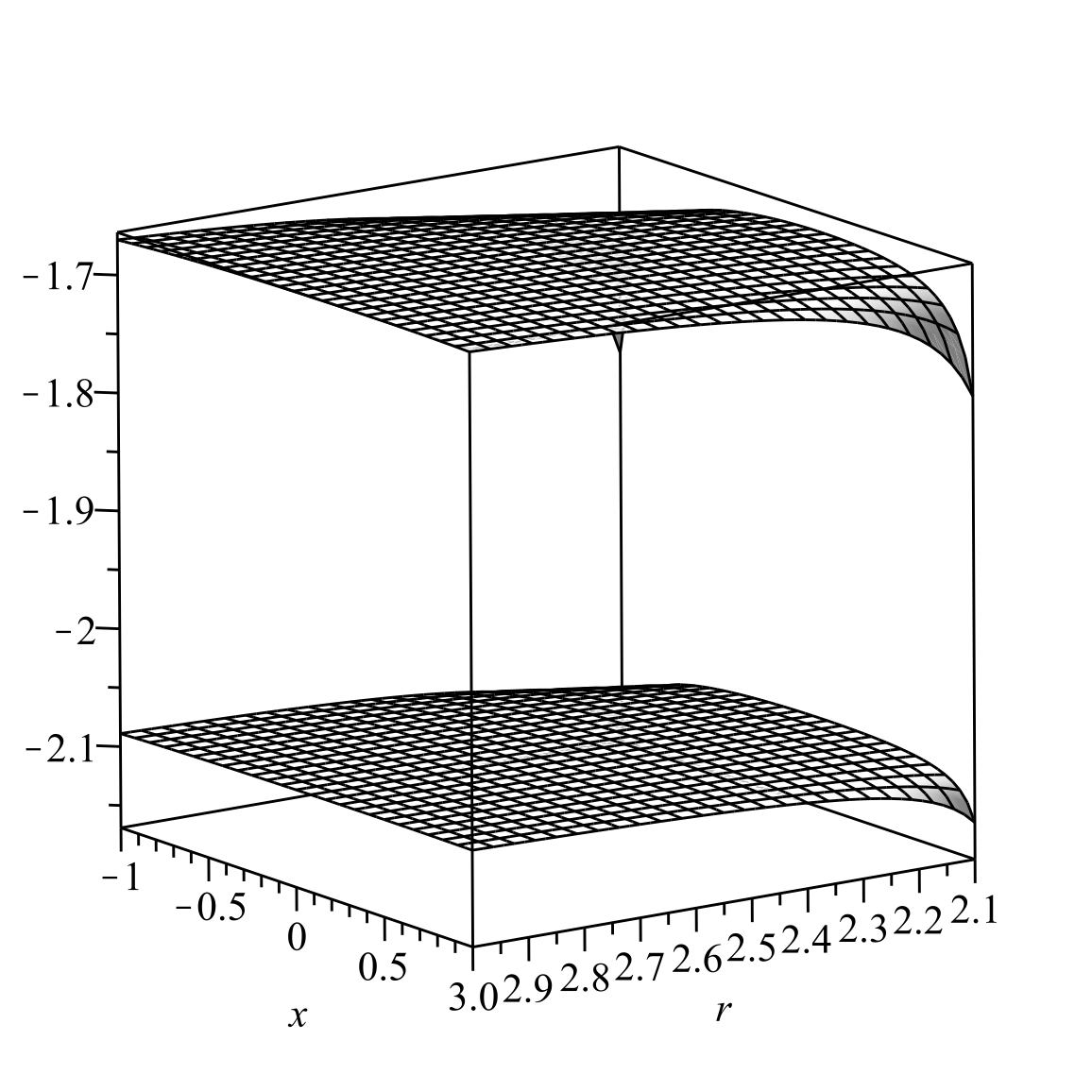}

\caption{The dilaton $\phi(t,r,\theta)$ as function of $r$ and $x=\cos\theta$  for two time slices $t=1$ (upper surface) and $t=2$ (lower surface), where we set $h=2,\,k_+=5,\,k_-=3,\,a=1,\mu=1,\nu=2$.}
\label{fig6}
\end{figure}

\begin{figure}[H]
\centering
\includegraphics[width=0.4\textwidth]{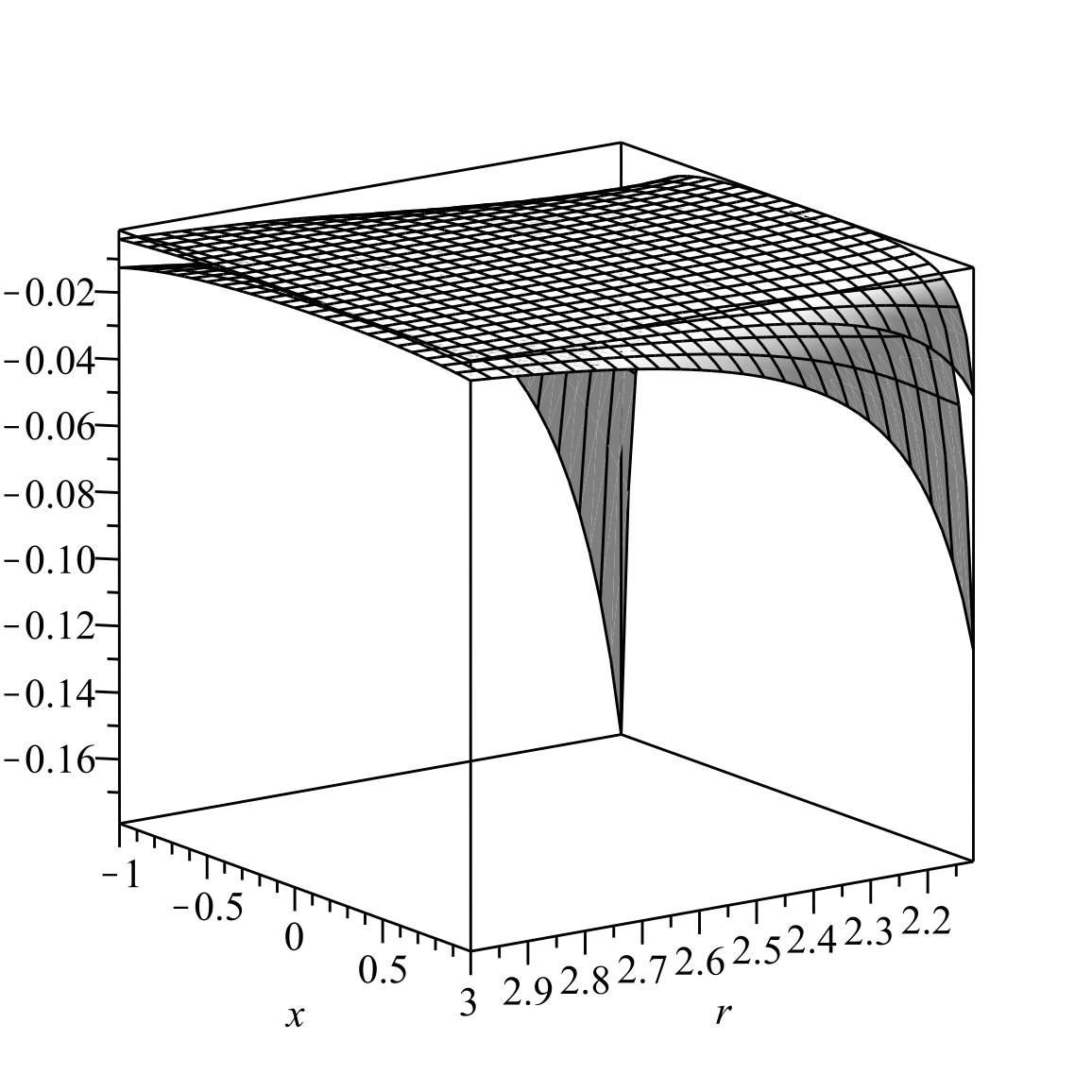}

\caption{The $r$-component of electric field 
as function of $r$ and $x=\cos\theta$ for two time slices; $t=1$ (lower surface), $t=2$ (upper surface), where we set $h=2,\,k_+=5,\,k_-=3,\,a=1,\mu=1,\nu=2$.}
\label{fig7}
\end{figure}

\begin{figure}[H]
\centering
\includegraphics[width=0.4\textwidth]{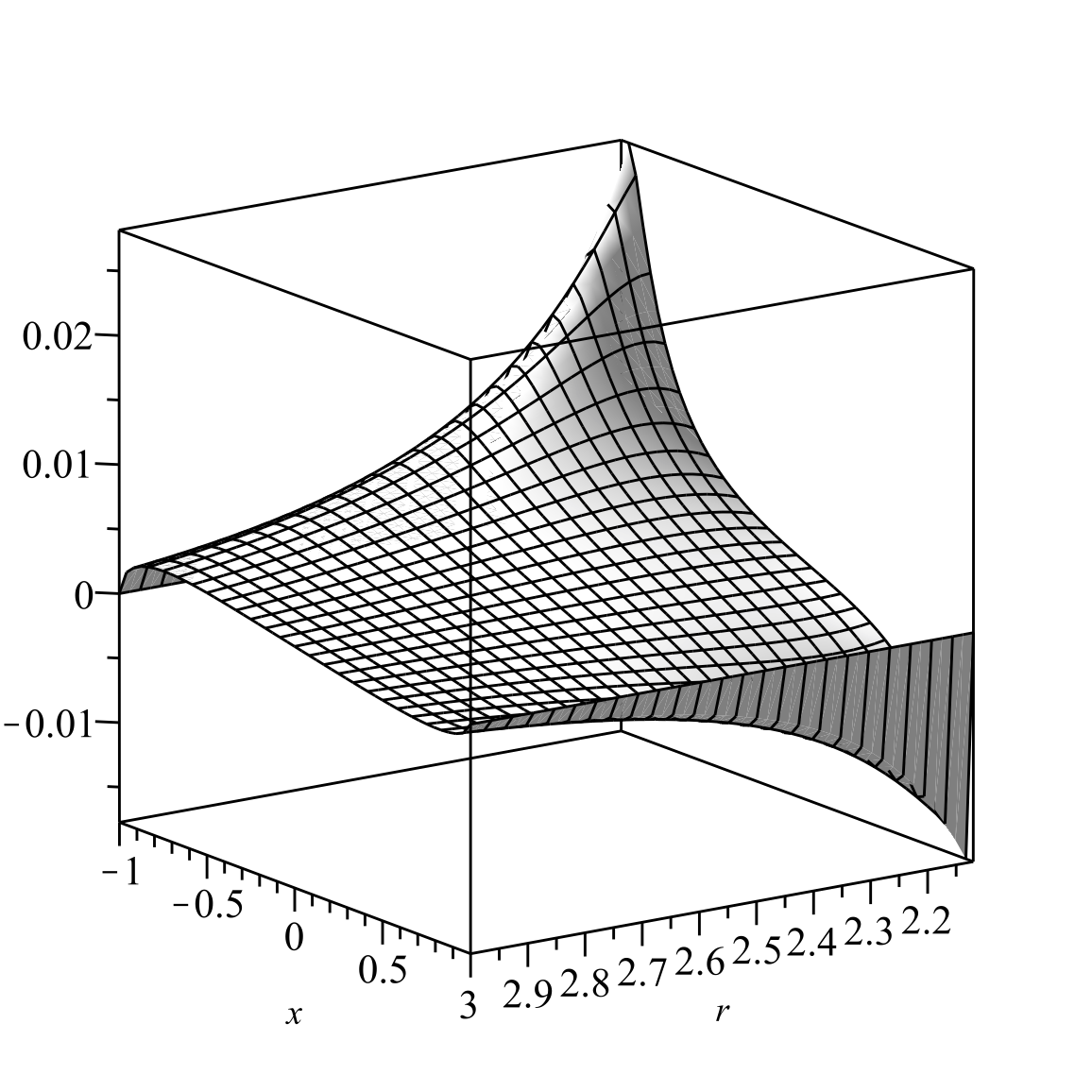}

\caption{The $\theta$-component of electric field 
as function of $r$ and $x=\cos\theta$ for one time slices; $t=1$, where we set $h=2,\,k_+=5,\,k_-=3,\,a=1,\mu=1,\nu=2$.}
\label{fig8}
\end{figure}

In section \ref{sec:aneqb}, we found a new class of exact solutions to Einstein-Maxwell-dilaton theory in which the two coupling constants are not equal and satisfy a constraint equation, as $ab=-2$. In this section, we find a new class of exact solutions to Einstein-Maxwell-dilaton theory, where the coupling constants $a$ and $b$ are equal. We consider the five-dimensional metric as
\begin{equation}
ds_5^{2}=-\frac{1}{H^2(t,r,\theta)}dt^{2}+R^2(t)H(t,r,\theta)ds_{EH}^2,
\label{dsaeqb}
\end{equation}
where $ds_{EH}^2$ is given by (\ref{EH}). The main difference between the metric ansatze  (\ref{dsanoteqb}) and (\ref{dsaeqb}) is that, in the former case, there is no time dependence for the metric function $H$, while in the latter case, the metric function $H$ depends explicitly on the time coordinate as well as two spatial coordinates.  We also use the following ansatz for the dependence of the electromagnetic field on the coordinates 
\be
{A_t}(t,r,\theta)={\alpha}{ R^X(t)H^Y(t,r,\theta)}\label{gaugeaeqb},
\ee
where $X$ and $Y$ are two constants that we will find them later.
Also, we consider a dilaton field that depends on both metric functions $H(t,r,\theta)$ and $R(t)$, and is given by
\be
\phi(t,r,\theta)=-\frac{3}{4a}\ln(R^U(t)H^V(t,r,\theta)),\label{phicase2}
\ee
where $U$ and $V$ are two other constants.  
Although we consider equal coupling constants, however solving the Maxwell's differential equations to find consistent values for the constants $U,V,X$ and $Y$ are more difficult, because the differential equations are more complicated (compared to case where the coupling constants are not equal), due to extra terms that are related to the time derivative of metric function $H(t,r,\theta)$. 
A detailed analysis of the Maxwell's equations and some of the Einstein field equations (appendix B) lead to 
\begin{eqnarray}
X&=&-a^2,\\
Y&=&-1-\frac{a^2}{2},\\
U&=&2a^2,\\
V&=&a^2,\label{cons}
\end{eqnarray}
and the following form for the metric function $H(t,r,\theta)$
\be
H(t,r,\theta)=R^{-2}(t)\{R^{2+a^2}(t)+K(r,\theta)\}^{\frac{2}{2+a^2}},\label{HGfixfromapC}
\ee
where $K(r,\theta)$ is an arbitrary function of $r$ and $\theta$. We substitute for $H(t,r,\theta)$ in terms of $R(t)$ and $K(r,\theta)$ in ${\cal G}_{tt}=0$ and symbolically solve to find the cosmological constant $\Lambda$.  We then substitute for the cosmological constant in ${\cal G}_{rr}=0$ and find the following differential equation for the metric function $R(t)$ 
\be
\left( R \left( t \right)  \right) ^{4\,{a}^{2}+2} \left( {\frac 
{\rm d}{{\rm d}t}}R \left( t \right)  \right) ^{2}({a}^{2}-1)
+ \left( R \left( t \right) 
 \right) ^{4\,{a}^{2}+3}{\frac {{\rm d}^{2}}{{\rm d}{t}^{2}}}R \left( 
t \right)=0,
\label{E11}
\ee
as well as two partial differential equations for $K(r,\theta)$, that are given by
\bea 
&-& 3\, r  \left( {
\frac {\partial ^{2}}{\partial {r}^{2}}}K \left( r,\theta \right) 
 \right) K \left( r,\theta \right) \sin \theta \, ({{
h}}^{4}-r^4)
+3 \left( {\frac {\partial 
}{\partial r}}K \left( r,\theta \right)  \right) K \left( r,\theta
 \right) \sin  \theta \, ({{h}}^{4}+3r^4)
 \nn\\
&+&12\, r^3 \left( {\frac {\partial }{\partial \theta}}K \left( r,
\theta \right)  \right) K \left( r,\theta \right) \cos  \theta
 +12 \, r^3 \left( {\frac {\partial ^{2}}{
\partial {\theta}^{2}}}K \left( r,\theta \right)  \right) K \left( r,
\theta \right) \sin \theta \nn\\
&+&4 \, r^3 \left( {\frac {\partial }{\partial 
\theta}}K \left( r,\theta \right)  \right) ^{2}\sin  \theta \,
  ({a}^{2}{\alpha}^{2}+2\alpha^2-3)=0,
  \nn\\
\label{E21}
\eea
and
\bea
&-&\, r \left( {\frac {\partial ^{2}}{
\partial {r}^{2}}}K \left( r,\theta \right)  \right) \sin 
\theta  \, (h^{4}-r^4)
+ \left( {\frac {
\partial }{\partial r}}K \left( r,\theta \right)  \right) \sin 
\theta \, (h^{4}+3\, r^4)\nn\\
&+&
 4\, r^3  \left( {\frac {\partial }{\partial 
\theta}}K \left( r,\theta \right)  \right) \cos \theta +4 \, r^3 \left( {\frac {\partial ^{2}}{\partial {\theta
}^{2}}}K \left( r,\theta \right)  \right) \sin  \theta  =0.\nn\\
&&
\label{E22}
\eea
We solve the differential equation (\ref{E11}) and find the solutions for $R(t)$ as
\be
R(t)=(\mu t+\nu)^{\frac{1}{a^2}}.\label{RR}
\ee
Moreover we find that the solutions to (\ref{E21}) and (\ref{E22}) are given by
\be
K(r,\theta)=1+\frac{k_+}{r^2+h^2\cos\theta}+\frac{k_-}{r^2-h^2\cos\theta},\label{GG1}
\ee
where $k_+$ and $k_-$ are two constants.
Quite interestingly, this function looks the same as the function $G(r,\theta)$, that we found before in (\ref{GG}). Substituting the metric functions $R(t)$ and $H(t,r,\theta)$ from equations (\ref{HGfixfromapC}), (\ref{RR}) and (\ref{GG1}) into the equation that we found for the cosmological constant, we find that the cosmological constant is 
\be
\Lambda=\frac{3}{2}\mu^2\frac{4-a^2}{a^4}.\label{CosmoC}
\ee
We note that the cosmological constant takes positive, zero or negative values, depending on the coupling constant $a$. After substituting all the obtained results in  the other remaining Einstein field equations, we find that all of them are completely satisfied. The dilaton field equation also is satisfied upon substituting the metric functions and the cosmological constant. 
Figures \ref{fig5}-\ref{fig8} show the typical behaviours of the metric function $H(t,r,\theta)$, the dilaton field $\phi(t,r,\theta)$ and the non-zero components of the electromagnetic tensor $F_{tr}(t,r,\theta)$ and $F_{t\theta}(t,r,\theta)$ versus coordinates $r$ and $\theta$, over some $t={\text {constant}}$ hypersurfaces, where we set the specific values for the constants $h=2,\,k_+=5,\,k_-=3,\,a=1,\mu=1,\nu=2$.

The five-dimensional metric 
\begin{eqnarray}
ds_5^2&=& -(\mu t+\nu)^{\frac{4}{a^2}}\{(\mu t+\nu)^{\frac{2+a^2}{a^2}}+\frac{k_+}{r^2+h^2\cos\theta}+\frac{k_-}{r^2-h^2\cos\theta}\}^{-4/(a^2+2)}dt^2\nn\\
&+&\{(\mu t+\nu)^{\frac{2+a^2}{a^2}}+\frac{k_+}{r^2+h^2\cos\theta}+\frac{k_-}{r^2-h^2\cos\theta}\}^{2/(a^2+2)}\nn\\
&\times&\{\frac{dr^2}{g(r)^2}+\frac{r^2g(r)^2}{4}(d\psi +\cos \theta d \phi)^2+\frac{r^2}{4}(d\theta^2+\sin^2(\theta)d\phi^2)\},\label{metraeqb}
\end{eqnarray}
definitely can be uplifted to a solution to the Einstein-Maxwell theory with a cosmological constant, in higher dimensions if the coupling constant $a$ is greater than or equal to  $\frac{1}{2}$ and less than $1$.  In fact, the Einstein-Maxwell theory with a cosmological constant, would be a $5+{\cal D}$-dimensional theory with the cosmological constant $\Lambda=\frac{9\mu^2}{2}\frac{({\cal D}+3)({\cal D}+4)}{{\cal D}^2}$, where ${\cal D}=\frac{3a^2}{1-a^2}$. The $5+{\cal D}$-dimensional metric is
\be
ds_{5+{\cal D}}^2=e^{\frac{4}{3}\sqrt\frac{{\cal D}}{{\cal D}+3}\phi(t,r,\theta)}ds_5^2+e^{\frac{-4}{{\cal D}}\sqrt\frac{{\cal D}}{{\cal D}+3}\phi(t,r,\theta)}d\vec y \cdot d\vec y,\label{highdEM}
\ee
where $\vec y=(y_1,\cdots ,y_{\cal D})$ parametrizes an Euclidean  $E^{\cal D}$ space. We explicitly check that (\ref{highdEM}) satisfies all the Einstein-Maxwell field equations for $a=\frac{1}{2},\sqrt{\frac{2}{5}}$ and $\frac{1}{\sqrt{2}}$ (where ${\cal D}=1,2$ and $3$) with the cosmological constants $\Lambda=90\mu^2,\,\frac{135}{4}\mu^2$ and $21\mu^2$, respectively. 
We also note that asymptotically, the metric (\ref{metraeqb}) becomes
\be
-dt^2+(\mu t)^{2/a^2}\{dr^{2}+\frac{r^2}{4}\big(d\Omega^2 +{(d\psi+\cos\theta d\phi)^2}\big)\}.\label{case2asym}
\ee
As we notice from (\ref{case2asym}), the $t=\text{constant}$ hypersurfaces are simply the fibration of a circle which is parameterized by $\psi$, over a sphere. We note that the scale factor for the $t=\text{constant}$ hypersurface is $(\mu t)^{2/a^2}$, while for the solutions with different $a$ and $b$, the scale factor of the asymptotic metric (\ref{rinf}) is $(\xi t+\zeta)^{a^2/2}$.  

We note that the solutions for the dilaton field (\ref{phicase2}) as well as the cosmological constant (\ref{CosmoC}), are not well defined where the coupling constant $a$, approaches to zero. In other words, if we are interested in the exact solutions with the finite cosmological constant, in the limit where $a\rightarrow 0$, we shall impose the limit in the action (\ref{act}) and solve the corresponding field equations. 


\section{Solutions with both coupling constants $a=b$ equal to zero}
\label{sec:aeq0}

\begin{figure}[H]
\centering
\includegraphics[width=0.4\textwidth]{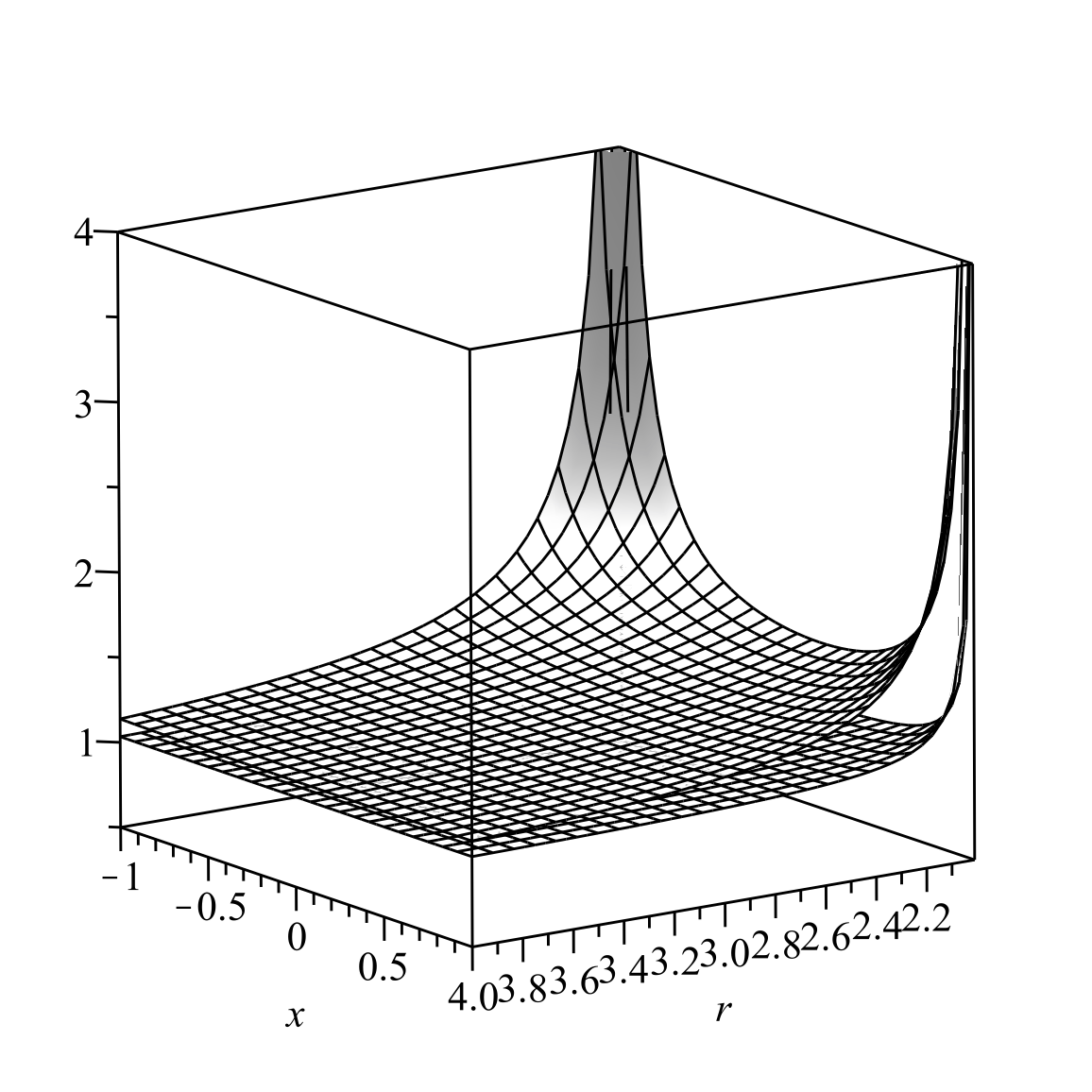}
\caption{The metric function $H(t,r,\theta)$  
as a function of $r$ and $x=\cos\theta$ for two time slices $t=1$ (upper surface) and $t=2$ (lower surface), where we set $\epsilon=+1,h=2,\,a_+=5,\,a_-=3,\,\Lambda=3$.}
\label{fig9}
\end{figure}

\begin{figure}[H]
\centering
\includegraphics[width=0.4\textwidth]{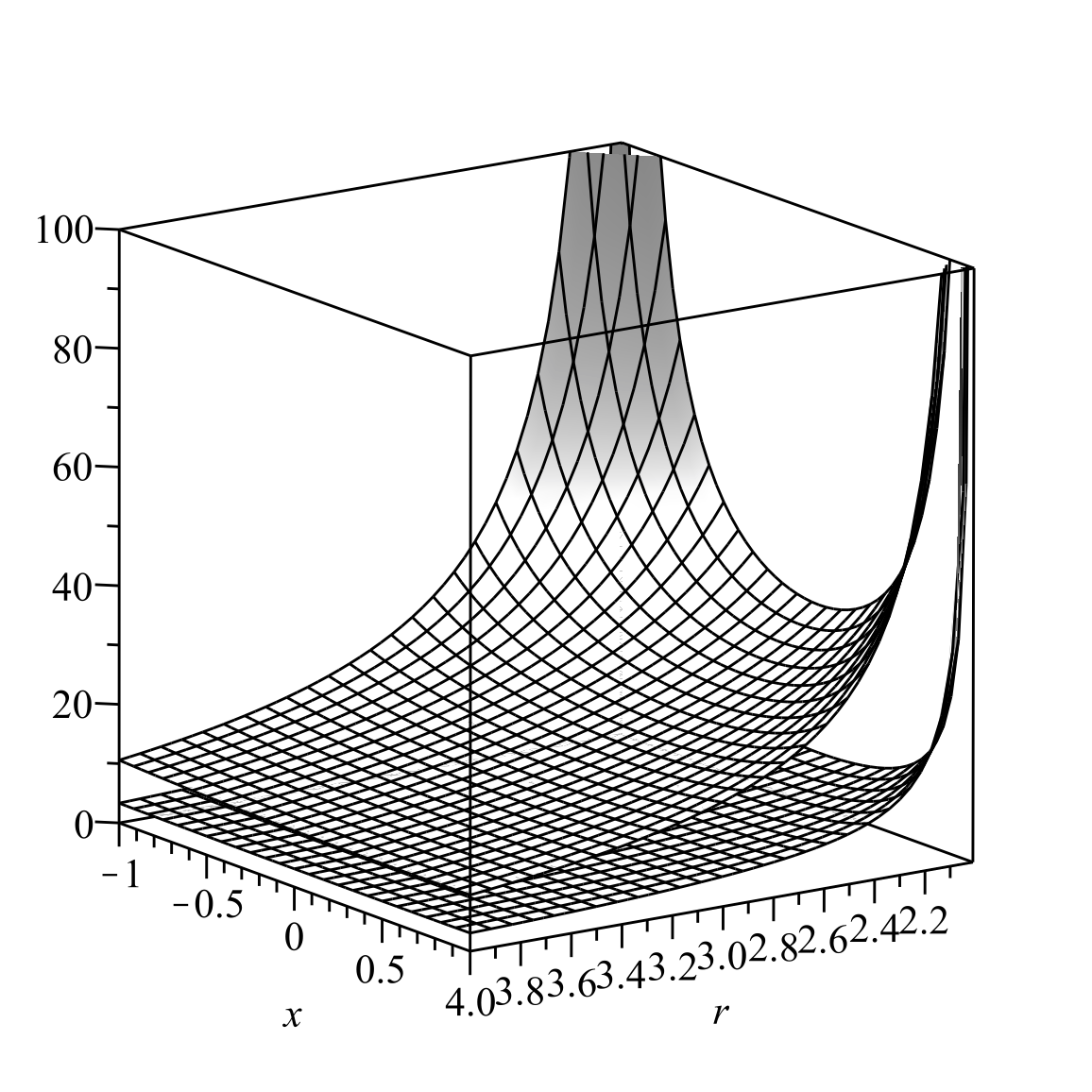}
\caption{The metric function $H(t,r,\theta)$  
as a function of $r$ and $x=\cos\theta$ for two time slices $t=1$ (lower surface) and $t=2$ (upper surface), where we set $\epsilon=-1,h=2,\,a_+=5,\,a_-=3,\,\Lambda=3$.}
\label{fig10}
\end{figure}

\begin{figure}[H]
\centering
\includegraphics[width=0.4\textwidth]{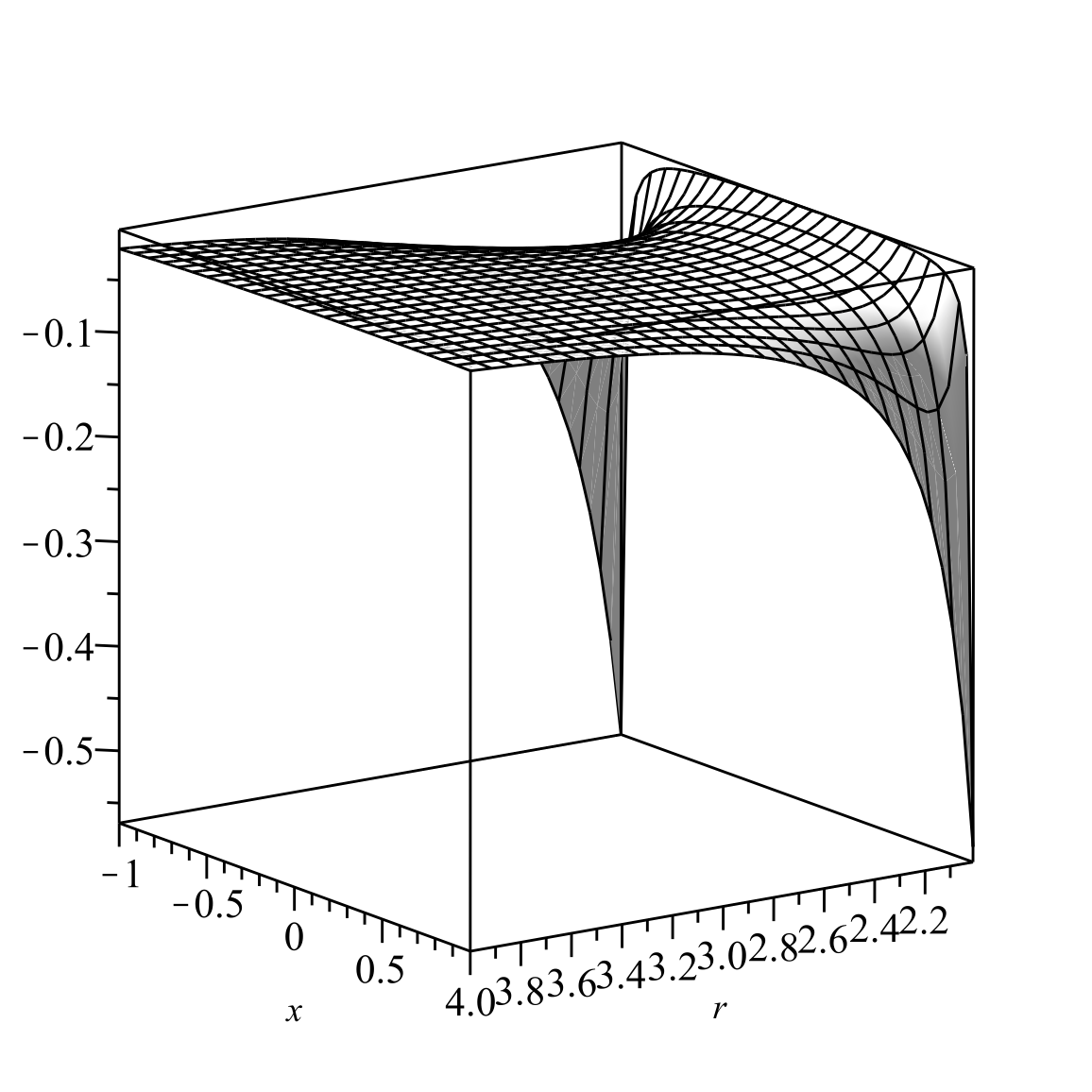}
\caption{The $r$-component of electric field 
as function of $r$ and $x=\cos\theta$ for $t=1$, where we set $h=2,\,a_+=5,\,a_-=3,\,\Lambda=3$.}
\label{fig11}
\end{figure}

\begin{figure}[H]
\centering
\includegraphics[width=0.4\textwidth]{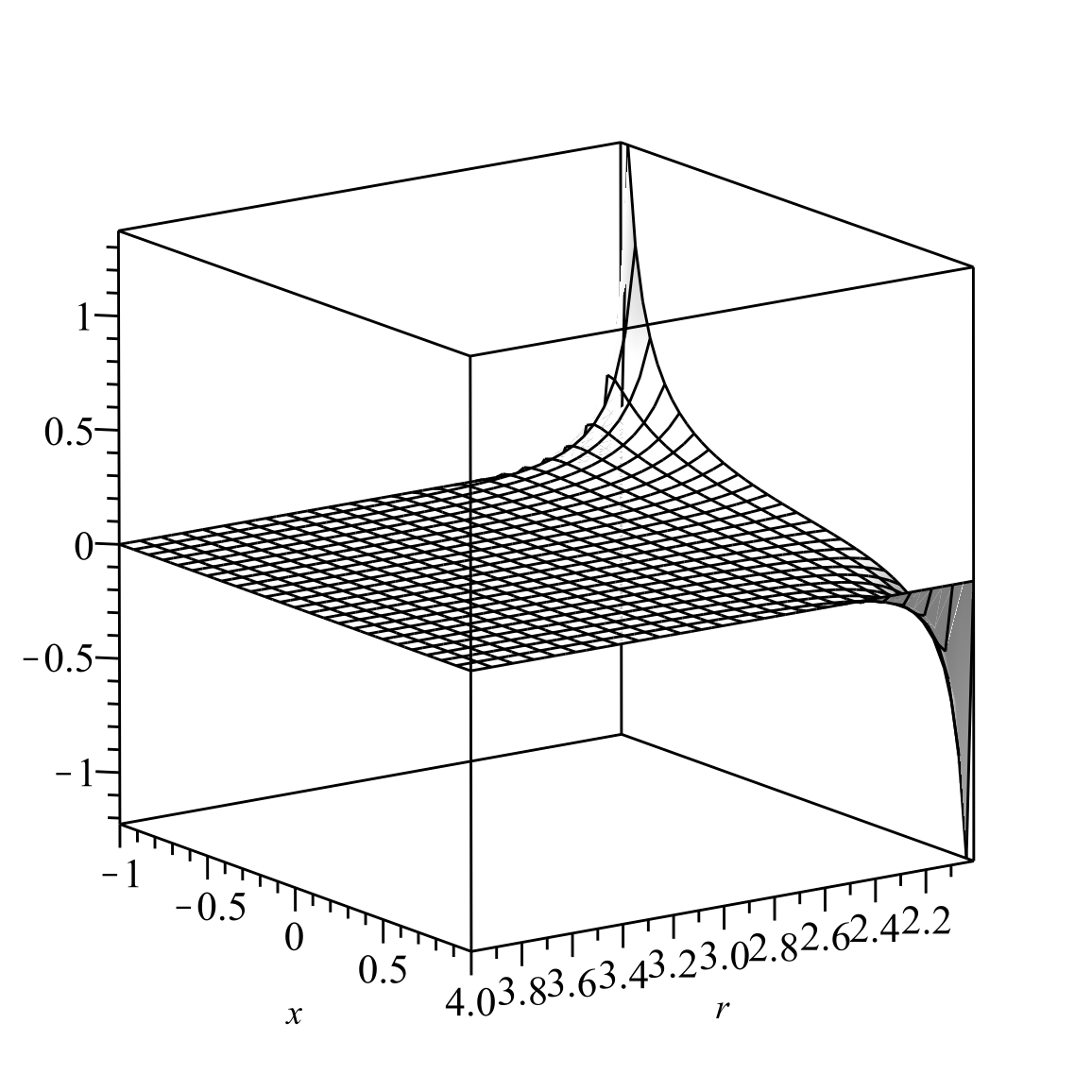}
\caption{The $\theta$-component of electric field 
as function of $r$ and $x=\cos\theta$ for $t=1$, where we set $h=2,\,a_+=5,\,a_-=3,\,\Lambda=3$.}
\label{fig12}
\end{figure}

As we noticed in the last section, we can't simply consider a solution where the coupling constant $a=b \rightarrow 0$, as this makes the dilaton field and cosmological constant divergent. In fact, in the limit of $a=b \rightarrow 0$, the action (\ref{act}) implies that the dilaton field decouples from the theory and the theory simply reduces to the Einstein-Maxwell theory in presence of a cosmological constant.  We consider the metric ansatz the same as (\ref{dsaeqb}) 
\begin{equation}
ds_5^{2}=-\frac{1}{H(t,r,\theta)^{2}}dt^{2}+R(t)^2H(t,r,\theta)ds_{EH}^2,
\label{dsaeqbzero}
\end{equation}
however, we consider a simpler ansatz for the Maxwell gauge field, that is given by
\be
{A_t}(t,r,\theta)=\frac{\alpha}{H(t,r,\theta)}\label{gaugeaeqbzero}.
\ee
Inspired with the solutions (\ref{HGfixfromapC}) and (\ref{GG1}) for the metric function $H$, where the coupling constants are not zero, we consider a possible solution for the metric function $H(t,r,\theta)$ as
\be
H(t,r,\theta)=1+\frac{f(t)}{r^2+X\cos\theta},\label{Hansatza0}
\ee
where $f(t)$ depends only on time and $X$ is a constant. Substituting the solution (\ref{Hansatza0}) in Maxwell's equation ${\cal M}^t=0$ leads to 
\be
X=\pm h^2.
\ee
So, we consider the following form for the metric function
\be
H(t,r,\theta)=1+{f(t)}(\frac{f_+}{r^2+h^2\cos\theta}+\frac{f_-}{r^2-h^2\cos\theta}),\label{Hansatzabzero}
\ee
where $f_\pm$ are two constants. The other Maxwell's equations ${\cal M}^r=0$ and ${\cal M}^\theta=0$, lead to a differential equation for the function $f(t)$,
\be
R(t)\frac{d f(t)}{dt}+2f(t)\frac{d R(t)}{dt}=0,
\ee
and so, we find that
\be
f(t)\simeq \frac{1}{R(t)^2}.
\ee
The non-diagonal field equations ${\cal G}_{tr}=0,\,{\cal G}_{t\theta}=0$ are satisfied upon substitution for $H(t,r,\theta)$ by equation (\ref{Hansatzabzero}). Moreover, the equation 
${\cal G}_{r\theta}=0$ yields
\be
\alpha^2=\frac{3}{2}.
\ee 
We then symbolically solve the equation ${\cal G}_{\phi\psi}=0$ for the cosmological constant $\Lambda$ and then substitute the result into equation ${\cal G}_{tt}=0$. We find a differential equation for the metric function  $R(t)$ as
\be
R(t)\frac{d^2 R(t)}{dt^2}-(\frac{d R(t)}{dt})^2=0,\label{Reqabzero}
\ee
where the solutions are given by
\be
R(t)=R_0e^{\epsilon R_1t}.\label{Rsolabzero}
\ee
In equation (\ref{Rsolabzero}), $R_0$ and $R_1$ are two constants and $\epsilon=\pm 1$.  The Einstein's equation ${\cal G}_{\phi\psi}=0$ leads to an algebraic equation for $R_1$
\be
6R_1^2=\Lambda.
\ee
To summarize, the metric function $H(t,r,\theta)$ finally is given by
\be
H(t,r,\theta)=1+e^{-2\epsilon \sqrt{\frac{\Lambda}{6}}t}(\frac{a_+}{r^2+h^2\cos\theta}+\frac{a_-}{r^2-h^2\cos\theta}),\label{Hansatzabzerofinal}
\ee
where $a_\pm$ are two arbitrary constants. We also explicitly check that the other Einstein's equations ${\cal G}_{\phi\phi}=0$, ${\cal G}_{rr}=0$ and ${\cal G}_{\theta\theta}=0$ are satisfied after substituting the above results.  Figures \ref{fig9} and \ref{fig10} show the typical behaviour of the metric function as a function of $r$ and $x=\cos\theta$ for $\epsilon=+1$ and $\epsilon=-1$ respectively. In the former case, the metric function monotonically decreases by increasing the time coordinate, while in the latter case, it monotonically increases by increasing the time coordinate. The two spikes in the figures correspond to the location of singularities at $r=h,\, \theta=0$ and $r=h,\, \theta=\pi$ that can be removed by a conformal rescaling. The components of electric field are plotted in terms of coordinates $r$ and $x=\cos\theta$ in figures \ref{fig11} and \ref{fig12}.  We should note that our solution for the metric function (\ref{Hansatzabzerofinal}) is different from the metric function for the coalescing black holes in Einstein-Maxwell theory with cosmological constant, on Eguchi-Hanson space \cite{CBH2, Co3}, due to the different metric ansatz (\ref{dsaeqbzero}).

\section{Conclusions}

In this article, we found explicit exact solutions to the five-dimensional Einstein-Maxwell-dilaton theory  in presence of cosmological constant where the dilaton field couples to the Maxwell fields as well as the cosmological constant. We first considered the most general case, where the two coupling constants are different.  We found that the constant number in the Maxwell field, the different exponents of the metric functions in the Maxwell and dilaton fields and the  cosmological constant depend on the value of the first coupling constant. Moreover the value of the second coupling constant is being fixed by the the first coupling constant. We derived the explicit form of the metric for the spacetime and showed that the solutions can't be uplifted to a higher dimensional Einstein-Maxwell theory or higher dimensional Einstein gravity with a cosmological constant. However a higher-dimensional gravity theory with a form-field without any initial cosmological constant can be compactified on an internal curved space to yield the uplifted version of our solutions. We discussed the asymptotic properties of the spacetime and showed that the spacetime is regular almost everywhere, up to a conformal rescaling.  We then considered the theory, where the two dilaton coupling constants are equal  and found explicit expressions for the metric, the Maxwell and dilaton fields. We discussed the physical properties of the asymptotic metric. We also found that for some special values of the dilaton coupling constant, we can uplift the solutions with the equal dilaton coupling constants to the solutions of a higher-dimensional Einstein-Maxwell theory where the dimension of theory depends on the dilaton coupling constant.  Moreover, we found a new class of exact solutions to the Einstein-Maxwell theory as a result of reduction of theory in the limit of zero dilaton coupling constant. The solutions are asymptotically dS while for non-zero dilaton coupling constants are not necessarily dS or anti dS.  Constructing similar solutions to the Einstein-Maxwell-dilaton theory with other self-dual geometries in five or higher dimensions would be an interesting task as well as studying the thermodynamics and relation of our solutions to the possible coalescing of black holes. We leave these open questions for a forthcoming article \cite{mme}. 

\section{Appendix A: The Field Equations for Einstein-Maxwell-Diltaon Theory}

The variation of the Einstein-Maxwell-dilaton action (\ref{act}) with respect to $g_{\mu\nu}$ leads to the Einstein's field equations that are given by
\begin{equation}
{\cal E}_{\mu\nu}\equiv R_{\mu\nu}-\frac{2}{3}\Lambda g_{\mu\nu}e^{4/3b\phi}-(F_{\mu}^{\lambda}F_{\nu\lambda}-\frac{1}{6}g_{\mu\nu}F^2)e^{-4/3a\phi}-\frac{4}{3}\nabla_\mu \phi \nabla _\nu \phi=0.\label{einstein}
\end{equation}

Moreover, varying the action (\ref{act}) with respect to the electromagnetic gauge field $A_\mu$ yields the Maxwell's field equations
\begin{equation}
{\cal M}_\mu\equiv\nabla ^\nu (e^{-4/3a\phi}F_{\mu\nu})=0,\label{maxwell},
\end{equation}
and finally, the variation of the action (\ref{act}) with respect to the dilaton field  gives
\begin{equation}
{\cal D}\equiv \nabla ^2 \phi + \frac{a}{4}e^{-4a\phi/3}F^2-be^{4/3b\phi}\Lambda=0\label{dilaton}.
\end{equation}

\section{Appendix B: Solving the Field Equations for Equal Coupling Constants $a=b$ }
The three non-zero Maxwell's equations after substituting for $\phi(t,r,\theta)$ (given by equation (\ref{phicase2})), are given by
\bea
{{\cal M}^t}{ }&=&3\,\alpha\, \sin\theta\, H^Y \left( t,r,\theta \right)   Y  R^{X}
 \left( t \right)  {r}^{5} \{ 
   \left( {\frac {\partial }{\partial r}}H \left( t,r,\theta
 \right)  \right) ^{2} (V\,{ h}^{4}r-V\,r^5+Yh^4r-Yr^5+h^4r-r^5)\nn\\
&+&
H \left( t,r,\theta \right)  \left( {\frac {\partial ^{2}}{
\partial {r}^{2}}}H \left( t,r,\theta \right)  \right) ({ h}^{4}
r-r^5)
 \nn\\
 &-&
  \left( {\frac {\partial }{\partial r}}H \left( t,r,
\theta \right)  \right) H \left( t,r,\theta \right)  ({ h}^{4}
-3r^4)
-4\, {r}^{3}
 \left( {\frac {\partial }{\partial \theta}}H \left( t,r,\theta
 \right)  \right) ^{2}( V+Y+1)\nn\\
\nn\\
&-&4\, \left( {\frac {\partial ^{2}}{\partial {\theta}^{2}}}H
 \left( t,r,\theta \right)  \right) H \left( t,r,\theta \right) {r}^{3}-4\,H \left( t,r,\theta \right)  \left( {
\frac {\partial }{\partial \theta}}H \left( t,r,\theta \right) 
 \right) {r}^{3}\tan ^{-1}  \theta  \},
\label{EQ1}
\\
{\cal M}^r&=&
\left( {\frac {\partial }{\partial r}}H \left( t,r,\theta \right) 
 \right)   \,\left( {\frac {\partial }{\partial t}}
H \left( t,r,\theta \right)\right) \,R \left( t \right)\,(V+Y+1) 
\nn\\
&+&
 \left( {\frac {\partial }{\partial r}}H \left( t,r,\theta \right) 
 \right) H \left( t,r,\theta \right) (U+X+2){\frac {\rm d}{{\rm d}t}}R
 \left( t \right) 
 +  
 \left( {\frac {\partial ^{2}}{\partial t\partial r
}}H \left( t,r,\theta \right)  \right) R \left( t \right) H \left( t,r
,\theta \right),
\label{EQ2}
\\
{\cal M}^\theta&=&
\left( {\frac {\partial }{\partial \theta}}H \left( t,r,\theta \right) 
 \right)   \,\left( {\frac {\partial }{\partial t}}
H \left( t,r,\theta \right)\right) \,R \left( t \right)\,(V+Y+1) 
\nn\\
&+&
 \left( {\frac {\partial }{\partial \theta}}H \left( t,r,\theta \right) 
 \right) H \left( t,r,\theta \right) (U+X+2){\frac {\rm d}{{\rm d}t}}R
 \left( t \right) 
 + 
  \left( {\frac {\partial ^{2}}{\partial t\partial \theta
}}H \left( t,r,\theta \right)  \right) R \left( t \right) H \left( t,r
,\theta \right).
\label{EQ3}
\eea

We solve equation (\ref{EQ1}) for $V$ and substitute the result in equations (\ref{EQ2}) and (\ref{EQ3}). We then solve these two equations to obtain $U$. We find the two solutions for $U$ are equal, if and only if
\be
\frac{\partial ^2 H}{\partial t\partial r}\frac{\partial H}{\partial r}=\frac{\partial ^2 H}{\partial t\partial \theta}\frac{\partial  H}{\partial \theta}.\label{cond}
\ee 
We find that equation (\ref{cond}) implies $\frac{\partial H}{\partial r}-\frac{\partial H}{\partial \theta}$ must be only a function of $r$ and $\theta$.  To satisfy the last requirement, we choose
\be
H(t,r,\theta)=f(t)\{g(t)+K(r,\theta)\}^W,
\ee
where $W$ is a constant and $f,g$ and $K$ are arbitrary functions.
Moreover, we notice that that the metric function $R^2H$ becomes
\be
R^2(t)H(t,r,\theta)=\{R^2(t)f(t)\}\{g(t)+k(r,theta)\}^W.
\ee
Hence, we may choose $f(t)=R^{-2}(t)$ and also $g(t)$ as some power of $R(t)$, to find possibly, the simplest solution for $H$. As a result, we consider the ansatz
\be
H(t,r,\theta)=R^{-2}(t)\{R^Q(t)+K(r,\theta)\}^W,\label{Hansatz}
\ee
where $Q$ is another constant.
We substitute (\ref{Hansatz}) for $H$ in Einstein's equation ${\cal G}^{tr}=0$ and  ${\cal G}^{t\theta}=0$. Both equations lead to 
\be
U=2V,\,W (V^2+2a^2)-2 a^2=0.\label{UW}
\ee
Moreover, substituting equation (\ref{Hansatz}) for $H$, as well as equation (\ref{UW}), in the Maxwell's equation ${\cal M}^{r}=0$ and also ${\cal M}^\theta=0$, yield the results
\be
Q(-2Va^2-2Ya^2+V^2-2a^2)+(X-2Y)(-V^2-2a^2)+2V^2+4a^2=0,\label{Mr1}
\ee
and
\be
(V^2+2a^2)(X-2Y-2)=0.\label{Mr2}
\ee
Equation (\ref{Mr2}) implies $Y=\frac{X}{2}-1$. We can't consider $V=-2a^2$ as this makes $W$, which is given by equations (\ref{UW}), diverges. If we consider the other non-diagonal Einstein equation ${\cal G}^{r\theta}=0$, we find that
\be
V+2Y+2=0\label{Grth1},
\ee
which implies $X=-V$ and also
\be
4\alpha^2Y^2a^2=3V^2+6a^2.
\ee
The last equation gives $\alpha^2$ in terms of $V$ as
\be
\alpha^2=\frac{3V^2+6a^2}{4(1+V/2)^2a^2}.
\ee
Substituting all the above obtained results in ${\cal G}^{r\theta}=0$, we find
\be
V=a^2.
\ee
The equations (\ref{UW}) give
\be
U=2a^2,\,W=\frac{2}{2+a^2}.
\ee
We also find from (\ref{Grth1}) that
\be
Y=-1-\frac{a^2}{2},
\ee
and $X=-a^2$. The constant Q also can be obtained from (\ref{Mr1}) that is given by
\be
Q=2+a^2.
\ee
So, we have found all the constants that appear in our solutions.  We summarize the results of this appendix, that
\be
A_t(t,r,\theta)=\frac{3}{a^2+2}R^{-a^2}(t)H^{-1-a^2/2}(t,r,\theta),\label{Afix}
\ee
\be
\phi(t,r,\theta)=\frac{-3}{4a}\ln(R^{2a^2}(t)H^{a^2}(t,r,\theta)),
\ee
and
\be
H(t,r,\theta)=R^{-2}(t)\{R^{2+a^2}(t)+K(r,\theta)\}^{\frac{2}{2+a^2}}.\label{HGfix}
\ee
 
 \vskip 1cm
{\Large Acknowledgments}

This work was supported by the Natural Sciences and Engineering Research
Council of Canada. The author would like to thank Rainer Dick for discussion.
\vskip 2cm


\end{document}